\documentclass[showpacs,preprintnumbers,amsmath,amssymb,floatfix]{revtex4-1}

\usepackage{color}      
\usepackage{graphicx}
\usepackage{subfig}
\usepackage{dcolumn}
\usepackage{bm}         

\usepackage{epsfig}

\begin{document}

\title{Optical model potential analysis of $\bar nA$ and $nA$ interactions}

\author{Teck-Ghee Lee$^{1}$ and Cheuk-Yin Wong$^2$}

\affiliation{$^1$Department of Physics, Auburn University, Auburn, AL
  36849, U.S.A.}

\affiliation{$^2$Physics Division, Oak Ridge National Laboratory, Oak
  Ridge, TN 37831, U.S.A.}
  
\begin{abstract}
We use a momentum-dependent optical model potential to analyze 
the annihilation cross sections of the antineutron $\bar n$ on C, Al, Fe, Cu, Ag, Sn, and Pb nuclei
for projectile momenta $p_{\rm lab}$ $\lesssim$ 500 MeV/$c$. We obtain a good description of 
annihilation cross section data of Barbina  {\it et al.}  [Nucl.~Phys.~A {\bf 612}, ~346~(1997)]  
and of Astrua {\it et al.} [Nucl.~Phys.~A {\bf 697},~209~(2002)] which exhibit an interesting dependence 
of the cross sections on $p_{\rm lab}$ as well as on the target mass number $A$. 
We also obtain the neutron ($n$) non-elastic reaction cross sections for the same targets. Comparing 
the $nA$ reaction cross sections $\sigma^{nA}_{\rm rec}$ to the $\bar nA$ annihilation cross sections 
$\sigma^{\bar nA}_{\rm ann}$, we find that $\sigma^{\bar nA}_{\rm ann}$ is significantly larger than 
the $\sigma^{nA}_{\rm rec}$, that is, the $\sigma^{\bar nA}_{\rm ann}$/$\sigma^{nA}_{\rm rec}$ 
cross section ratio lies between the values of about 1.5 to 4.0 in the momentum region where comparison is possible. 
The dependence of the $\bar n$ annihilation cross section on the projectile charge is also examined 
in comparison with the antiproton $\bar p$. Here we predict the $\bar pA$ annihilation cross section 
on the simplest assumption that both $\bar pA$ and $\bar nA$ interactions have the same nuclear 
part of the optical potential but differs only in the electrostatic Coulomb interaction. 
Deviation from a such simple model extrapolation in measurements will provide 
new information on the difference between $\bar nA$ and $\bar pA$ potentials. 

\end{abstract}


\maketitle

\section{Introduction}
Annihilation between an antinucleon and a nucleon or nucleus defines 
one of the basic aspects in antimatter-matter interactions. Over the years 
there have been many experimental measurements \cite{Arm87,Ber97,H1,H2,H3,H4,FI,NA,
Bia11,Bia00a,Bia00b,D2,D3,He1,He2,He3,C,Ne, Al, Kle05, asacusa18} and 
theoretical studies \cite{Kle05,Lee14,Lee16, Mah88,Kuz94, Car93,Car97,
Bia00,Gal00,Bat01,Fri14, Fri15, Gal02,Gal03,Gal08,Gal11,Gal11PL,
Gal12,Gal13,Gal14,Gal15} about antinucleon annihilation on nucleons 
and nuclei. However, most of the work was carried out with the
antiproton $\bar p$ projectile. Experimental and theoretical investigations 
using the antineutron $\bar n$, on the other hand, are still relatively limited.
Theoretical work has also been carried out on the relationship between $\bar nn$ oscillation 
and the $\bar nA$ interaction potential \cite {dover83, Kondratyuk,Gal}.   
Recently, it has also been suggested that $\bar nA$ annihilation
can be used to prepare an apparatus for $\bar nn$ 
oscillations \cite{nnbarOsc} detection.  

On the experimental side, one representative investigation is the measurement of the 
$\bar n$-Fe annihilation cross section from 100 to 780 MeV/$c$ \cite{Agnello88, Barbina97, Ableev94}. 
The experiment was carried out with the LEAR facility at CERN using the $\bar p p$ $\rightarrow$ $\bar n n$ 
charge-exchange reaction. Another  investigation, by the OBELIX group of Astrua {\it et al} \cite{NA}, measured the 
annihilation cross section of $\bar n$ on C, Al, Cu, Ag, Sn, and Pb nuclei in 
the $p_{\rm lab}$ range from 50 to 400 MeV/$c$. These experiments give clear evidence 
about the dependence of the antinucleon-nucleus absorption cross section on the mass 
number $A$ and about the momentum dependence which exhibits the prominent absorption feature 
of inverse-momentum dependence at low-energies. They are also useful to test the theories 
of antinucleon-nucleus interactions.    

In response to the experimental efforts, Friedman 
derived an optical model potential for $\bar p$-nucleus interaction by accounting for both the 
neutron and proton densities \cite{Fri15} to examine the annihilation cross sections for $\bar p$ 
and $\bar n$ on all the six targets at seven energies studied in Astrua {\it et al} \cite{NA}. 
The calculated cross sections for $\bar p$ and $\bar n$ 
were compared with experimental annihilation cross sections for $\bar n$. 
The study indicated that the $\bar p$ induced annihilation cross sections increase 
much more steeply in the low momentum  $p_{\rm lab} <$ 200 MeV/$c$ region in comparison to 
the case for the $\bar n$ projectile. It also elucidated that the larger $\bar p$ annihilation cross sections 
match the experimental data closely, but surprisingly not for $\bar n$ annihilation 
cross sections. Above 250 MeV/$c$, the $\bar n$ annihilation cross sections are found to be reasonably 
close to the experimental cross sections. However, below 100 MeV/$c$, the cross sections
 are found to be significantly smaller than the experimental cross sections. 
Furthermore, the predicted $\bar n$ annihilation cross sections display the feature of decreasing 
and shifting to lower and lower momenta as the size of the nuclear target increases 
and thus deviate from the behavior suggested by the experimental cross sections. It is important to 
note that the very same density-folded optical model potential was checked and 
tested previously, by the same author of Ref.\cite{Fri15}, to reproduce very well the 
angular distributions for elastic scattering of $\bar p$ by C, Ca and Pb at 
300 MeV/$c$ \cite{Fri14}. 

The fact that $\bar n$ induced annihilation cross sections
are smaller than for $\bar p$ can be easily understood  because the incoming electrically 
neutral projectile will naturally experience negligible Coulomb attraction from the target nucleus. 
But, it is perplexing that, experimentally there is a notable absorption feature of $1/p_{\rm lab}^\alpha$-like 
dependence, akin to the effects of Coulomb focusing for $\bar n$ annihilation cross sections
at the lower momenta, and the microscopic optical potential predicted that these cross sections 
decrease and shift to lower and lower momenta as $A$ increases.

Recently, we have extended the Glauber model for nucleus-nucleus collisions 
\cite{Gla59, Gla70, Won84,Won94} to study the nuclear annihilation cross sections 
by antinucleons. The extended Glauber model for the calculation 
of the $\bar p A$ annihilation cross section \cite{Lee14, Lee16} considered 
the nucleon-nucleus collision as a collection of binary collisions, and took into account
the appropriate shadowing and the inclusion of initial-state and in-medium interactions. 
The basic ingredients are the elementary $\bar p p$ and $\bar p n$ annihilation 
cross sections, $\sigma_{\rm ann}^{\bar p p}$ and $\sigma_{\rm ann}^{\bar p n}$, 
together with initial-state Coulomb interactions and the change of the momentum of 
the antinucleon inside the nuclear medium. We note that in our 
earlier study \cite{Lee14}, the basic $\bar p p$ annihilation cross section, 
$\sigma_{\rm ann}^{\bar p p}$, was parametrized semi-empirically 
as $1/v$, and employed in our investigation of the stability and the properties of matter-antimatter 
molecules \cite{Lee11, Lee08}. In our subsequent study \cite{Lee16}, 
we improved the $\sigma_{\rm ann}^{\bar p p}$ and $\sigma_{\rm ann}^{\bar p n}$ 
formulas by considering the anti-particle transmission through a 
nuclear potential and the $\bar p p$ Coulomb interaction, thereby the nuclear 
annihilation cross sections can be properly evaluated in a simple analytical form. The
expressions are rigorous enough and therefore we amend our
earlier simple approach of a $1/v$ function to parametrize the 
basic $ \sigma_{\rm ann}^{{\bar p p}}$ and $ \sigma_{\rm ann}^{{\bar p n}}$ 
cross sections. The strong absorption model formulated decomposes the
incoming plane waves into a sum of partial waves of given orbital
angular momentum $L$ and assumes that these partial waves transmitted to the
nucleon surface $S$ lead to an annihilation reaction.  It is shown that the
cross sections for nuclear annihilation by $\bar p$ and $\bar n$ are
simple functions of the momentum of the incident particles. Across the
momentum range considered, contrasting it to the $ \sigma_{\rm
  ann}^{{\bar n p}}$ annihilation cross section, the $ \sigma_{\rm
  ann}^{{\bar p p}}$ annihilation cross section is significantly
enhanced by the Coulomb interaction for the $p_{\rm lab}$ momenta of
the incident particle below 500 MeV/$c$. As the $p_{\rm lab}$
increases, the two annihilation cross sections become almost
identical, approaching the Pomeranchuk's equality limit \cite{Pom56} at $p_{\rm
  lab}$ $\sim$ 500 MeV/$c$. In addition, the calculated annihilation
cross sections agree well with the experimental data. With the improved 
$ \sigma_{\rm ann}^{{\bar p p}}$ and $ \sigma_{\rm ann}^{{\bar p n}}$, 
we also reproduced the general map of annihilation cross sections,  
$ \sigma_{\rm ann}^{{\bar p A}}$, as a function of nuclear mass 
numbers $A$ and collision energies. 

With encouraging results from the particle transmission theory to describe the 
$ \sigma_{\rm ann}^{{\bar p p}}$, $ \sigma_{\rm ann}^{{\bar p n}}$ and
$ \sigma_{\rm ann}^{{\bar p A}}$ annihilation cross sections, we employed 
the very same theory to examine the $ \sigma_{\rm ann}^{{\bar n A}}$. 
But there was an inadvertent error that arose through the Coulomb trajectory modification 
considered in the extended Glauber model, making our $ \sigma_{\rm ann}^{{\bar n A}}$ 
to agree with the experiment data. We re-examined and re-evaluated our $\sigma_{\rm ann}^{{\bar n A}}$ 
cross section, and found, in the absence of additional Coulomb effects, that the rectified $ \sigma_{\rm ann}^{{\bar n A}}$ 
cross sections are significantly ``flat" and relatively lower than the experimental data for $p_{\rm lab}$ $<$ 200 MeV/$c$, 
yielding a far from satisfactory agreement between our calculations and experiment of Astrua {\it et al}.

Anticipating that new and better experiments \cite{FAIR09, Pan13, Mau99} will be performed in the coming years, here 
we attempt to explore an alternative theoretical method to rectify our previous 
annihilation cross section results for $\bar nA$. Moreover, it appears that a comparative 
study of the absorption cross sections induced by neutrons, antineutrons and antiprotons 
has not yet been made. 

The content of this paper is as follows. In Section II, we present 
the phenomenological optical model potential (OMP) we obtained to examine 
the $\bar n A$ annihilation cross sections. In Section III, we assess our 
phenomenological theory by comparing our numerical results to the available 
experimental 
$\bar n A$ annihilation cross section, $n A$ reaction cross section, and $\bar p A$ annihilation 
data. Finally, we conclude the present study with some discussions 
in Section VI.

\section{Phenomenological momentum-dependent optical model potential}

The Glauber model is known to work best at high energies in 
which the extend individual nucleon can be treated as an isolated scatterer.
For low-energy collisions, such a description may not be as appropriate,  
and the traditional optical model potential analysis may be more suitable.
For this reason we adopt a phenomenological analysis to study the 
energy-dependence of the 
OMP on $\bar n A$ 
annihilation cross section.  
Moreover, the method of OMP is well-tested and long-established 
for treating complicated interactions between
an incoming nucleon and a nucleus \cite{Varner91, Koning03}.

In the present analysis, we consider the collision between an 
antinucleon and a nucleus, and their effective interaction strength without 
spin-orbit interaction  is represented generally by a momentum-dependent optical model potential
\begin{eqnarray}
U(r) = V_C(r) - V_V(r,p) -i(W_V(r,p) + W_D(r,p)),
\end{eqnarray}
where subscripts ``$V$" and ``$D$" denote the volume and surface terms, respectively; and  
\begin{eqnarray}
V_V(r,p) &=& V_o(p) f(r, r_V, a_V),   \\
W_V(r,p) &=& W_o(p) f(r, r_W, a_W), \\
W_D(r,p) &=& -4a_{W_D}W_{o_D}(p) \frac{d}{dr}f(r, r_{W_D}, a_{W_D}). 
\end{eqnarray}
As usual the $f(r, r_x, a_x)$ is a Wood-Saxon form factor
\begin{eqnarray}
f(r, r_x, a_x) = \frac{1}{(1+\exp[(r-r_x)/a_x])},
\end{eqnarray}
where $x~\equiv~V, W, W_D$. The Coulomb term $V_C(r)$ is naturally zero for an 
electrically neutral projectile. Otherwise, 
\begin{eqnarray}
V_C(r) = 
    \begin{cases}
\frac{Z_A Z_p e^2}{2r_c}\left(3-\frac{r^2}{r^2_c}\right) & \quad \text{for~} r \leq r_c, \\ 
\frac{Z_A Z_p e^2}{r}~~~~~~~~~~~~~~~~~~~~~~~~& \quad \text{for~} r  > r_c, \\
    \end{cases} 
\end{eqnarray}
for a charged projectile with $Z_A$ and $Z_p$ being the target and projectile nuclear charges, 
respectively, and $r_c = r_oA^{1/3}$ is the Coulomb radius with $r_o$ being 1.25 fm. 

Although the main focus here is the $\bar nA$ optical model potential,
our knowledge of the $\bar pA$ optical model potential is more extensive. To gain 
some intuitions about the shape and size of our desired OMP, knowledge 
of the $\bar p A$ OMP is valuable as it could shed some light on the construction 
of $\bar n A$ OMP. There are at least two families of the $\bar p A$ 
potential, and these families and their ambiguity were studied by 
one of the present authors in \cite{Wong84}. One family, so-called $S$, 
has a much more shallow imaginary potential with $W$ of order 15$-$45 MeV, 
associated with a deep real potential with $V$ of order 200$-$350 MeV. 
The other one, so-called $D$, has a real well-depth $V$ of order 100 MeV 
and a deep imaginary part $W$ of order 100$-$200 MeV. On the other hand, 
the neutron-nucleus ($n A$) optical potential is also well-established. 
From Koning and Delaroche \cite{Koning03}, we learned that the $n A$ 
optical potential has a real well-depth $V$ of the order of 60 MeV and 
considerably shallower imaginary potential with $W$ of the order of 15 MeV 
for many nuclei across the periodic table, but with $A$-value greater than 23. 
This potential family is quite different from that of $\bar p A$. 

The optical model potential of Koning and Delaroche has many advantages 
because of its simplicities and systematic variations. However, as it has not taken 
into account the effects of static and dynamical deformation of the nuclei, it has 
its limitations and its application to $^{12}$C as we do here will exhibit an expected 
deficiency. 

It is desirable to have a simple, ``flexible'', and yet rich enough ({\it i.e.,} applicable 
in the very low momentum region) forms of optical model potential for $\bar nA$ that could also 
be useful for $\bar pA$ annihilation. We therefore concocted a momentum-dependent 
phenomenological optical model potential \begin{eqnarray}
V_o(p_{\rm lab}) = V'_o\times\left(\frac{\cosh(\sqrt{(b_0+p_{\rm lab})}-\sqrt{b_0})}{\cosh(\sqrt{(b_1+p_{\rm lab})}-\sqrt{b_1})} \right),   
\label{POMP}                       
\end{eqnarray}
where $b_0 $ and $b_1$ are two adjustable parameters. We choose this form so that
$V_o$ $\rightarrow$ $V'_o$ as $p_{\rm lab}$ $\rightarrow$ 0, and we use the $\rm cosh$ 
function such that $V_o(p_{\rm lab})$ decreases monotonically and gradually with $p_{\rm lab}$. In addition,
we also want our $V_o(p_{\rm lab})$ to behave similarly to the functional dependence of $V_V(E)$ 
of Koning and Delaroche  plotted in Fig.\ 1 of Ref.\cite{Koning03}. We also assume that 
our absorptive potentials, $W_o(p) = W_o$ and  $W_{oD}(p) = W_{oD}$, do not  vary with the projectile momentum.
Table~\ref{t1} lists the optical model potential well depths and the $b_0 $ and $b_1$  
parameters used in the calculations.

\begin{table} [h]
\caption{Antineutron optical model potential well depths, $V'_o$ and $W_{(o,oD)}$, 
and the $b_{(0,1)}$ free parameters, are in MeV, and $V_D = 0$. The $W_{(o,oD)}$ parameters 
are independent of the projectile momentum.}
\begin{tabular}{cccccccc}
\hline
\hline \\
Nucleus &  $^{12}$C & $^{27}$Al &  $^{56}$Fe & $^{63.6}$Cu & $^{107.9}$Ag & $^{118.7}$Sn & $^{206}$Pb \\
\hline
\hline \\
  $V'_o$  &  52.00 & 66.00 &  56.00 & 60.00 &  82.00 &  90.00 & 110.00   \\
  $W_o$   & 12.00 &  3.50 & 9.00  & 4.33  &  4.10 & 4.30 & 2.80  \\
  $W_{oD}$ & 5.98 & 5.98 & 5.98 & 5.98 & 5.98 & 5.98 & 5.98 \\
   $b_0$  & 14.04  &  31.86  &  67.20 & 75.52  & 127.29  & 140.08 & 243.08    \\ 
   $b_1$  & 7.92  &   16.90  &  39.00 & 37.70  & 61.10  & 65.00 & 106.60    \\ 
\\
\hline
\end{tabular}
\label{t1}
\end{table}

With regard to the radius parameter in the optical potential, we use
the following procedure to estimate its approximate value before more
refined search and adjustment.  From the experimental annihilation cross section at
high energies at which a geometrical approximation is a reasonable
assumption, we estimate a radius $r_R$ given by 
\begin{eqnarray}
\sigma_{\rm ann} = \pi (r_R A^{1/3})^2.  
\end{eqnarray} 
This radius defines a sharp cut-off distribution for the collision process.  The
equivalent Wood-Saxon optical model potential with a radius parameter
of $r_V$ and a diffuseness $a_V$ can be estimated by \cite{Hasse}

\begin{eqnarray}
r_V = r_R\left(1-\frac{1}{3}\left(\frac{\pi a_V}{r_RA^{1/3}}\right)^2\right) 
\label{rv}
\end{eqnarray} 
for each nuclei. For example, even though the $\bar n$C experimental annihilation cross section  
at $p_{\rm lab} > $ 500 MeV/c is not readily available, according to Pomeranchuk's 
equality at the high-energy limit \cite{Pom56}, both the $\bar n$C and $\bar p$C annihilation cross sections 
should be identical. Therefore, it is reasonable to make use of the experimental data to determine the value 
of $\bar p$C annihilation cross section at 900 MeV/$c$ and use this value to determine  
the $r_R$, which turns out to be 1.653 fm. Concerning how one guesses the value of the diffuseness parameter $a_V$, 
its initial estimate is deduced from the clues given by Friedman \cite{Fri14}, in which the $a_V$ 
for antineutron may be about a factor of 2-3 times of that for the neutron. To search for the optimal 
value of $a_V$, several iterative calculations for annihilation cross section have to be performed at 
a fixed momentum of 900 MeV/$c$ for both the $\bar n$C and $\bar p$C until both their
annihilation cross sections closely satisfy the Pomeranchuk's equality. 
Once the $r_R$ and $a_V$ values are determined, 
Eq.(\ref{rv}) gives the corresponding value of $r_V$. 
The same procedure is also applied to the case of iron nuclei. 

\begin{table} [htp]
\caption{The estimated annihilation cross sections at 400 MeV/$c$ and 900 MeV/$c$, 
and their corresponding values of $r_R$.}
\begin{tabular}{|c|c|c|c|}
\hline
Pair            & $p_{\rm lab}$(MeV/$c$) & $\sigma=\pi (r_RA^{1/3})^2$(fm$^2$) & $r_R$(fm)   \\
\hline
$\bar p$C  &        900                           &          45.0                              &     1.653       \\
$\bar n$Al  &       400                           &          100.0                             &    1.881      \\
$\bar p$Fe  &       900                          &           100.0                            &     1.475     \\ 
$\bar n$Cu  &       400                         &            180.0                          &      1.893     \\
$\bar n$Ag  &      400                          &             240.0                          &       1.840      \\ 
$\bar n$Sn  &      400                          &             265.0                          &       1.868      \\
$\bar n$Pb  &       400                          &            400.0                          &        1.911            
\\
\hline
\end{tabular}
\label{t2}
\end{table}

With respect to the Al, Cu, Sn and Pb nuclei (e.g., see Fig.\ 5 in Ref.\cite{Bia11}), 
despite the fact that there are $\bar p$ experimental data are available at around 1 GeV/$c$, 
they were not measured at a common momentum point. As a result, we are afraid that they 
can complicate the consistency of our estimations for the $r_R$ and hence $r_V$ 
values for each element. To be safe, we choose to use the experimental $\bar nA$ 
annihilation cross section values at 375 MeV/$c$ and extrapolate them 
to 400 MeV/$c$. Note that the same iterative $a_V$-search procedure is 
also considered for these elements. Table~\ref{t2} presents the annihilation cross sections at 400 MeV/$c$ and 900 MeV/$c$, 
and their corresponding values of $r_R$. The subsequent antineutron radial and diffuseness parameters 
for the POMP as a function of mass numbers are given in Table~\ref{t3}. Fig. \ref{fig8}(a) illustrates the variation 
of the strength of $V_o$ as a function of mass numbers and antineutron momentum. 
In general, their behaviors bear similarity with the momentum functional form of $\sigma^{\bar nA}_{\rm ann}$.  

\begin{table} [ht]
\caption{Optical model potential parameters for $\bar n A$ and $nA$ interactions. 
The neutron optical model potential parameters are from Ref.\cite{Koning03}. The geometry 
parameters $r_x$ and diffusiveness parameters $a_x$ are in fm. It is assumed 
that $r_W$ = $r_V$, $a_W = a_V$, $a_{V_D} = a_{W_D}$ and $ V_D = 0$.}
\begin{tabular}{ccccccccc}
\hline
\hline \\
 Nucleus & & $^{12}$C & $^{27}$Al &  $^{56}$Fe & $^{64}$Cu & $^{108}$Ag & $^{119}$Sn & $^{206}$Pb \\  [0.5ex] 
\hline\hline \\
   &$r_V$    & 1.234 & 1.577 & 1.307 & 1.649 & 1.663 & 1.681 & 1.785 \\ 
  &$r_{W_D}$ & 1.260 & 1.260 & 1.260 & 1.260 & 1.260  & 1.260 & 1.260 \\
 \raisebox{1.5ex}{$\bar n$} & $a_V$ & 1.050 & 1.250 & 1.050 & 1.500 & 1.500 &  1.600 &  1.600  \\
  &$a_{W_D}$ &0.590 &0.590 &0.590 &0.590 &0.590 &0.590 &0.590 \\ \\
\hline\\
   &$r_V$    & 1.127 & 1.162 & 1.186 & 1.203  & 1.219  &  1.221 & 1.235   \\ 
  &$r_{W_D}$ & 1.306 & 1.290 & 1.282 & 1.279 & 1.267 & 1.264 & 1.249 \\
 \raisebox{1.5ex}{$n$} & $a_V$ & 0.676 & 0.665 & 0.663 & 0.668 & 0.662  & 0.660 & 0.647 \\
  &$a_{W_D}$ & 0.543 & 0.538  & 0.532  & 0.534 & 0.527 & 0.525 & 0.510 \\ \\
\hline
\hline
\end{tabular}
\label{t3}
\end{table}

In order to obtain the $nA$ reaction cross section, we adopted the optical model potential
by Koning and Delaroche \cite{Koning03}. To avoid later confusion, we shall use the 
phenomenological optical model potential (POMP) to denote the antinucleon-nucleus 
interactions $U(r)$ of eq.(\ref{POMP}). On the other hand, we shall use the 
Koning-Delaroche's optical model potential (KD-OMP) to denote the $nA$ 
optical potential described in Ref.\cite{Koning03} . 

These optical model potentials are then employed in the Schr\"{o}dinger equation, and 
the standard distorted wave method provided in the ECIS97 computer program \cite{Raynal97} 
is used to solve the Schr\"{o}dinger equation to obtain the reaction cross section.  For each individual nucleus, 
we use a fixed value for $V_o$ evaluated at $p_{\rm lab}$ = 200 MeV/$c$ for 
$p_{\rm lab} \ge$ 200 MeV/$c$, as $V_o$ becomes almost constant in the high-energy limit. 
Furthermore, we also check the sensitivity of the cross section at $p_{\rm lab}$ = 200 MeV/$c$ 
with respect to the small variation ($\sim5\%$) of $V_o$ and make sure that the changes in the cross section 
is not more than $\sim5\%$. 

\section{Results and Discussion}

In this section, we first evaluate our $\bar nA$ annihilation cross section results 
by comparing with the available experimental data. Second, we discuss the differences between the $\bar nA$ annihilation 
and $nA$ reaction cross sections, and compare their corresponding optical model potential parameters. 
Third, we consider the $\bar pA$ annihilation. Lastly, we analyze the power laws 
of the $\bar p$ and $\bar n$ annihilation cross sections. 

\subsection{$\bar nA$ annihilation cross sections}

\begin{figure}[!htp]
\begin{center}
\includegraphics[scale=0.325]{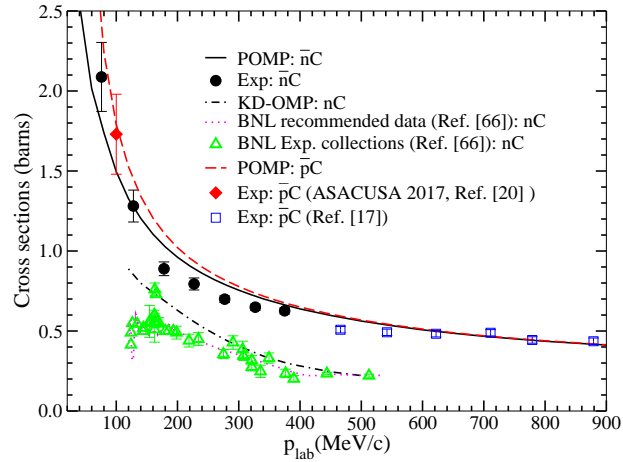}
\caption{(Color online)  Comparison of $\bar n$C,  $\bar p$C annihilation cross
sections and the $n$C non-elastic reaction cross section as a function of the projectile 
momentum in the laboratory frame. The dash-dot-dotted line
refers to the $n$C reaction cross section obtained using the KD-OMP; 
the dotted line and the scattered triangles are the $n$C reaction 
non-elastic data from Brookhaven National Laboratory's National 
Nuclear Data Center \cite{BNL}.}
\label{fig1}
\end{center}
\end{figure}

In our previous study \cite{Lee16}, we examined the $\bar n p$ annihilation 
cross section as a function of the antineutron 
momentum by considering the transmission through a nuclear potential. Although 
the  annihilation cross section data for $\bar n p$ still remain rather sparse to date 
in comparison to $\bar p p$ and contain significant degrees of uncertainty, a good 
agreement is achieved between our analytical results and experimental data from the OBELIX 
Collaboration \cite{Ber97} and from Brookhaven National Laboratory \cite{Arm87}. Similarly, 
a good way to verify and validate the present optical model potential model 
in describing the mass $A$ and momentum dependencies of $\bar n$ 
annihilation (and of $n$ reactions) is to benchmark against the 
available experimental data. 

\begin{figure}[!htp]
\begin{center}
\includegraphics[scale=0.325]{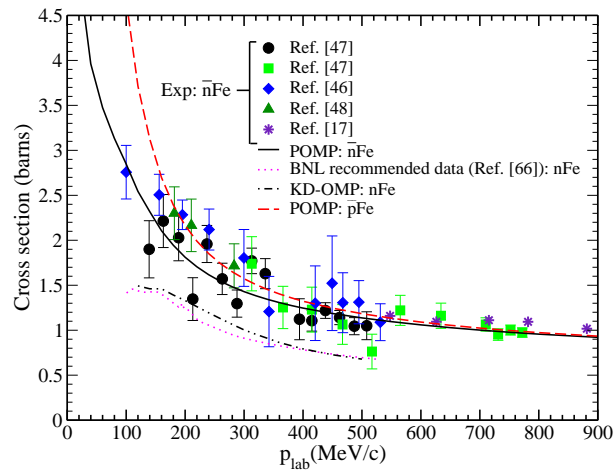}
\caption{(Color online) Comparison of $\bar n$Fe, $\bar p$Fe annihilation cross
  sections and the $n$Fe non-elastic reaction cross section as a function of the projectile 
  momentum in the laboratory frame. The symbols are experimental data 
  of $\bar n$Fe annihilation. The dash-dot-dotted line
refers to the $n$Fe reaction cross section obtained using the KD-OMP; 
  the dotted line is the $n$Fe reaction non-elastic data from 
  Brookhaven National Laboratory's National 
Nuclear Data Center \cite{BNL}.}
\label{fig2}
\end{center}
\end{figure}  

Fig.\ 1 shows a comparison of $\bar n$C annihilation cross sections against 
several sets of data. From a quantitative perspective,  the predicted cross sections appear to obey the 
momentum dependence behavior suggested by the experiment in the low-momenta region. 
As $p_{{\rm lab}}$ proceeds to increase beyond 500 MeV/$c$, the theoretical and experimental 
cross sections continue to remain in agreement, indicating that the annihilation cross section decreases 
as momentum increases. 

\begin{figure}[!tbp]
\begin{center}
\subfloat[Cross section.]{\includegraphics[width=8cm, height=12cm]{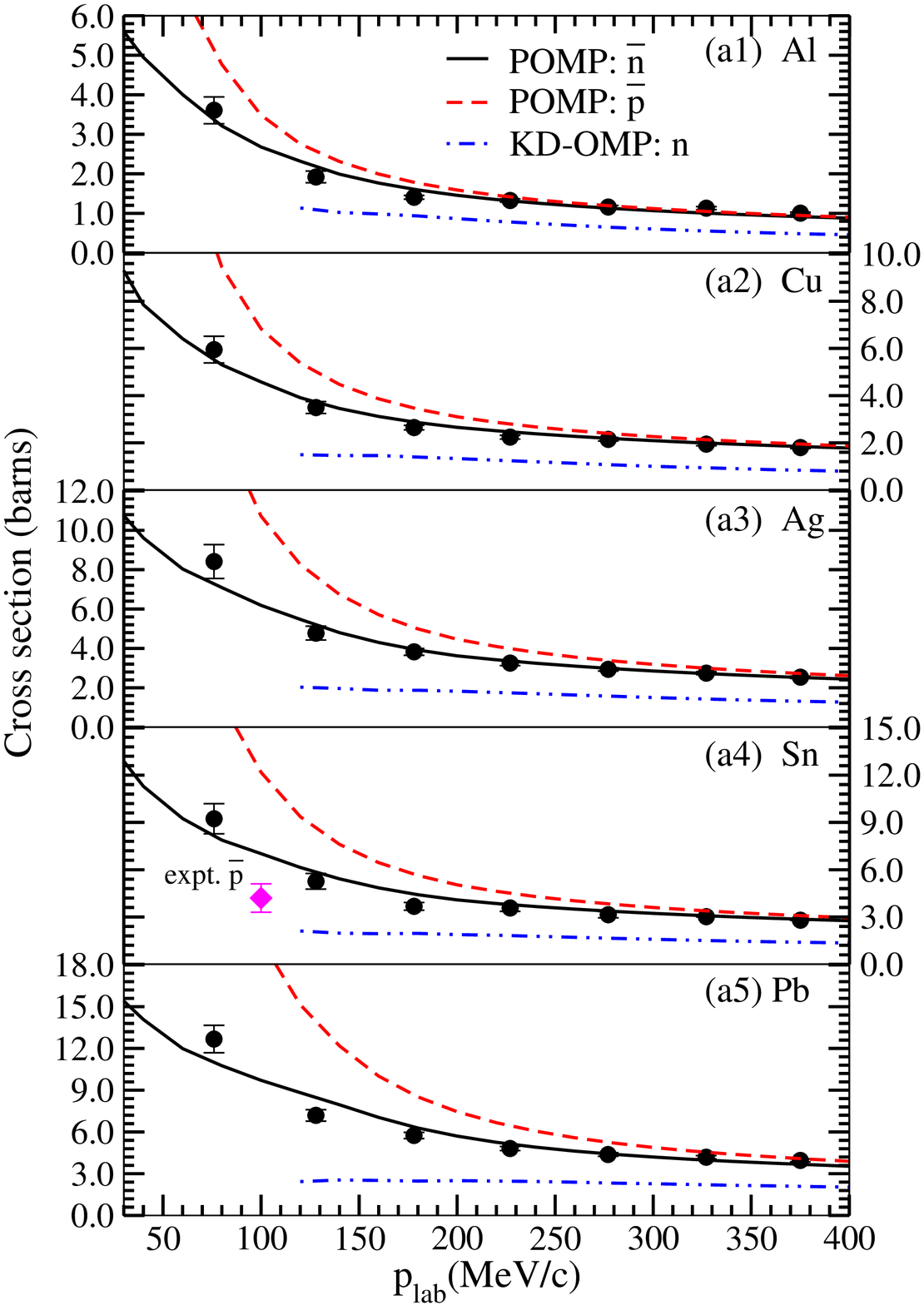}\label{fig:f1}}
  \hspace*{-0.77cm}
\subfloat[Cross section ratio.]{\includegraphics[width=8cm, height=12cm]{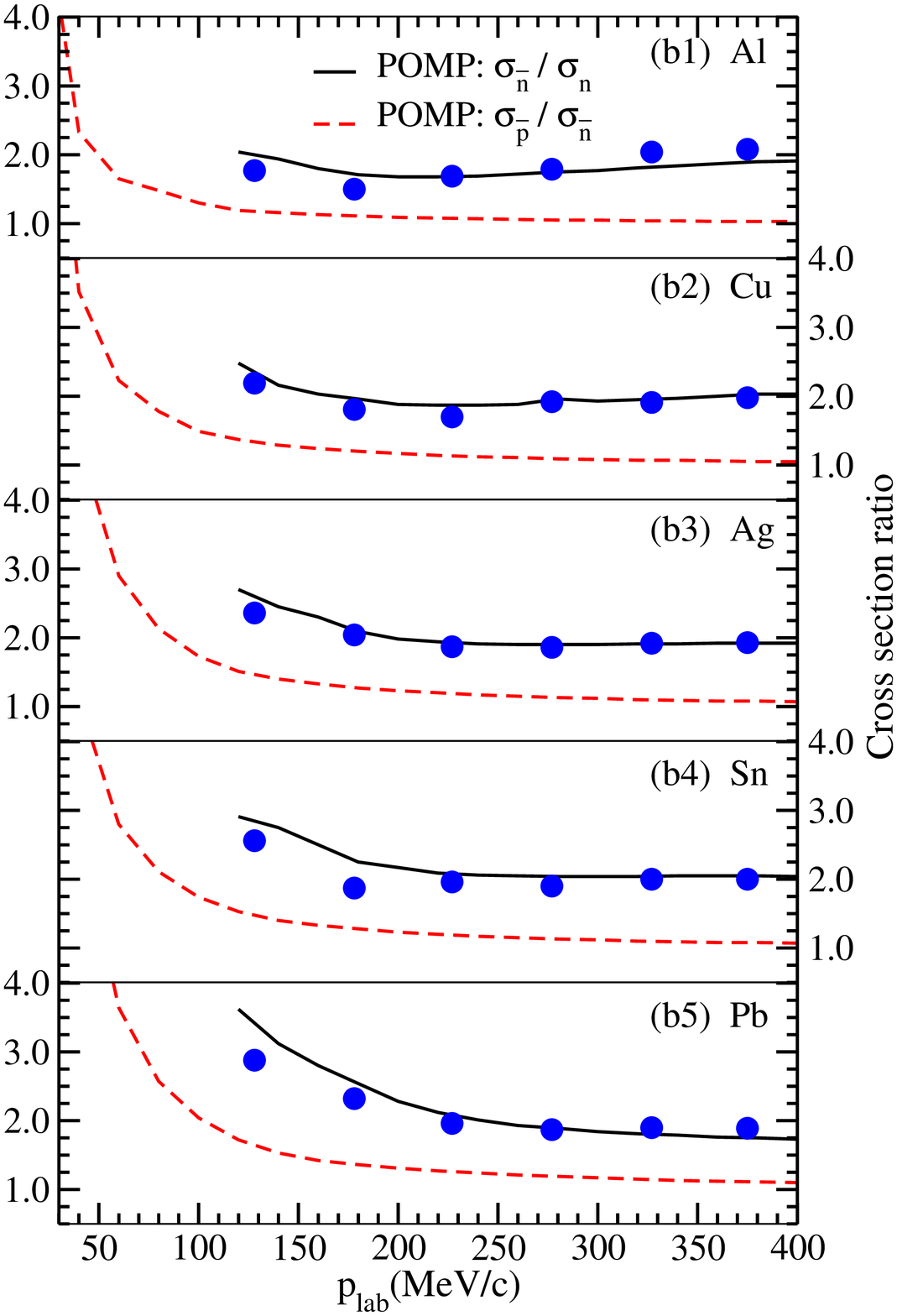}\label{fig:f2}}
\caption{(Color online)(a) Cross
  sections for $\bar nA$, $\bar pA$, and $nA$, as a function of the projectile momentum in the laboratory
  frame. The solid line is for $\bar nA$, the dashed line is for $\bar pA$, and the dash-dot-dotted line is for $nA$. 
  The circles are experimental data from Astrua {\it et al} \cite{NA}. The diamond is from Bianconi {\it et al.} \cite{Bia11}. 
  (b) Cross section ratios. The solid line represents the theoretical $\sigma^{\bar nA}_{\rm ann}$/$\sigma^{nA}_{\rm rec}$, 
  the dashed line represents the theoretical $\sigma^{\bar pA}_{\rm ann}$/$\sigma^{\bar nA}_{\rm ann}$ and 
  the solid circle represents the ratio of the experimental $\sigma^{\bar nA}_{\rm ann}$ to the theoretical $\sigma^{nA}_{\rm rec}$.}
\label{fig4}
\end{center}   
\end{figure}

In Fig.\ 2, we examine the $\bar n$Fe annihilation cross sections along 
with several data sets. Similarly, the calculated $\bar n$Fe annihilation cross 
sections also appeared to be in good agreement with the experimental data which 
indicate a much larger cross sections (in comparison to the case of $\bar n$C 
annihilation) below $p_{\rm lab}$ of 400 MeV/$c$ and the absorption feature becomes
progressively smaller as one goes up in $p_{\rm lab}$.

Fig.~\ref{fig4}(a) shows that the predicted $\bar nA$ annihilation cross sections 
for the Al, Cu, Ag, Sn and Pb nuclei rise considerably as the projectile momentum 
continues to decrease. These theoretical cross sections also describe the experimental 
data \cite{NA} relatively well in the momenta region where the data are available for comparison, 
except at $p_{{\rm lab}}$ of 76 MeV/$c$ where the calculations underestimated the
experiment by about 15-20$\%$ for Ag, Sn and Pb targets. In regard to the 
finding of Ref.\cite{Fri15} where $\bar n$ annihilation cross sections shift 
to lower and lower momentum as nuclear size increases, inspecting the change 
of $\sigma^{\bar nA}_{\rm ann}$ cross sections as a function the nuclear mass 
number $A$ displayed in Fig. \ref{fig4}(a), we do not notice any sign of 
reduction of $\bar nA$ annihilation cross sections and shift of such kind. 

\subsection{$nA$ reaction cross sections}
\begin{figure}[!htp]
\begin{center}
\includegraphics[width=10cm, height=8cm]{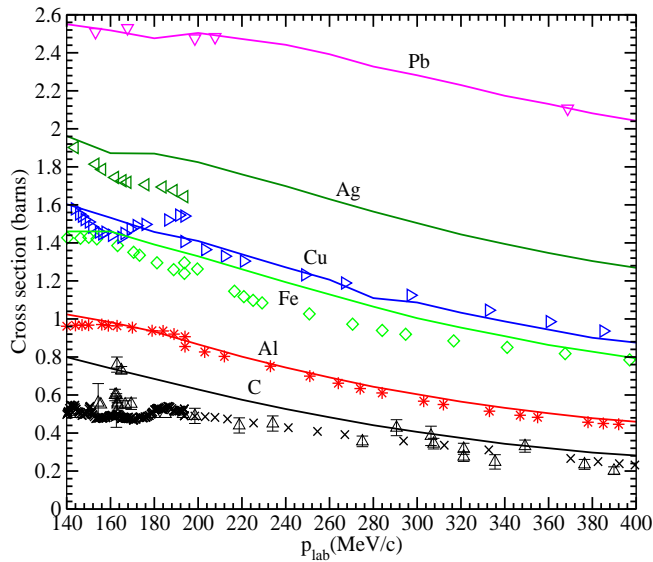}
\caption{(Color online) The $nA$ non-elastic reaction cross section as a function of the projectile 
  momentum in the laboratory frame. The solid-line refers to the present results obtained using 
  the KD-OMP. The symbols are the data recommended by the 
  Brookhaven National Laboratory's National 
Nuclear Data Center \cite{BNL}.}
\label{fig7}
\end{center}
\end{figure}  

The energy dependence of $nA$ reaction cross sections have been relatively well studied for many elements 
across the periodic table over the years. Therefore, it is meaningful to compare the $\bar nA$ annihilation cross section 
against  the $nA$ reaction cross section as a function of incoming projectile momentum. But before we 
do that, it is worthwhile to examine the quality of the present neutron reaction cross sections 
based on the KD-OMP. Displayed in Fig. \ref{fig7} is a comparison between the present 
results and the BNL recommended non-elastic reaction cross section data for C, Al, 
Fe, Cu, Ag, and Pb nuclei \cite{BNL}. It is shown that the overall agreement between 
the calculated cross sections and recommended data is reasonably good. 
Note that we intentionally left out the Sn results in the plot because, to our best knowledge, 
we could not find the available BNL data to make a comparison.

\begin{figure}[!tbp]
\begin{center}
\includegraphics[scale=0.5]{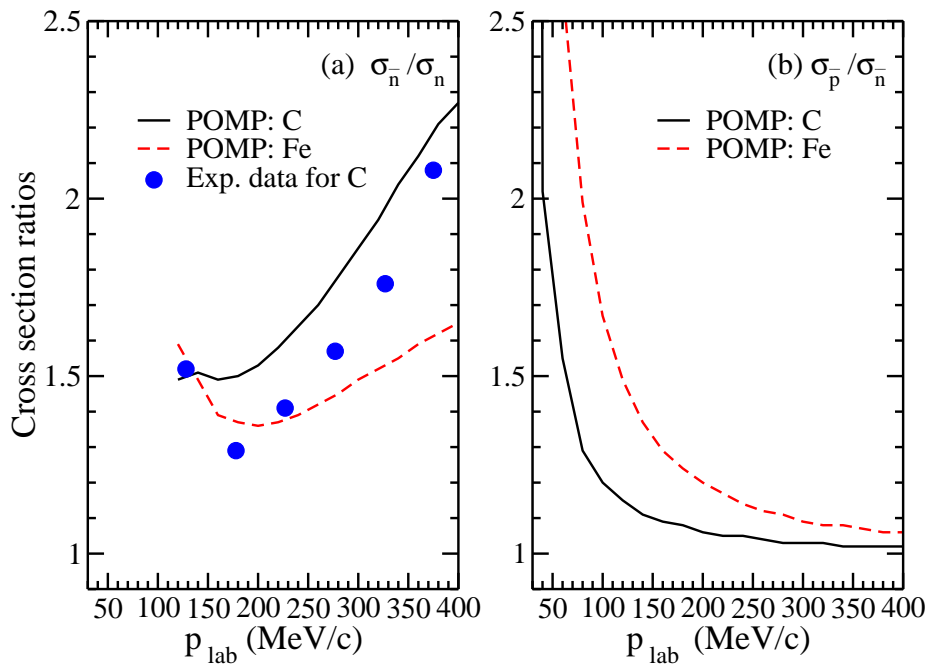}
\caption{(Color online) Cross section ratios as a function of the projectile momentum in the laboratory frame. 
(a) The ratios of $\sigma_{\bar n}/\sigma_{n}$. The solid line represents the theoretical
$\sigma^{\bar n \rm C}_{\rm ann}/\sigma^{n \rm C}_{\rm rec}$, the dashed line 
represents the theoretical $\sigma^{\bar n \rm Fe}_{\rm ann}/\sigma^{n \rm Fe}_{\rm rec}$, and the solid circle represents
the ratio of experimental $\sigma^{\bar n \rm C}_{\rm ann}$ to the theoretical $\sigma^{n \rm C}_{\rm rec}$. 
Note that there are no experimental $\sigma^{\bar n \rm Fe}_{\rm ann}$ available at the common momentum 
points of carbon. (b) The ratios of $\sigma_{\bar p}/\sigma_{\bar n}$. The solid line represents the 
theoretical $\sigma^{\bar p \rm C}_{\rm ann}/\sigma^{\bar n \rm C}_{\rm ann}$ and the dashed line represents 
the theoretical $\sigma^{\bar p \rm Fe}_{\rm ann}/\sigma^{\bar n \rm Fe}_{\rm ann}$.}
\label{fig3}
\end{center}
\end{figure}  

Since both the $\bar nA$ and $nA$ interactions are free from initial-state Coulomb interactions, it is valuable 
to compare the momentum dependence of the cross sections of these two interactions. One can clearly see, 
from figures \ref{fig1}, \ref{fig2}, \ref{fig4}(a), that the $\bar n$ annihilation cross sections 
of all targets are significantly larger than that of the $n$ reaction. To better appreciate their 
differences in the cross sections between the $\bar n$ and $n$ projectiles, 
we plot the $\sigma^{\bar nA}_{\rm ann}$/$\sigma^{nA}_{\rm rec}$ ratios 
as a function of the projectile momentum for carbon and iron nuclei in Fig.~\ref{fig3}(a).  
In general, the curves for carbon and iron nuclei depicted a similar behavior. In the same plot, 
we also include the ratio of experimental $\sigma^{\bar n \rm C}_{\rm ann}$ to the 
theoretical $\sigma^{n \rm C}_{\rm rec}$, which we shall denote as  the experimental ratio. 
It is interesting to see the shape of the curve of experimental ratios also resembles the behavior 
of the theory even though the agreement between the theoretical predicted and the experimental ratios 
is not that satisfactory. The disagreement may be attributed to the calculation not taking 
into account the effects of static and dynamical deformation of the carbon nuclei. 

Examining Fig.~\ref{fig3}(a) more closely, one finds that the theoretical 
$\sigma^{\bar n \rm C}_{\rm ann}$/$\sigma^{ n \rm C}_{\rm rec}$ 
ratio is about 1.5 at $p_{\rm lab}$ $\simeq$ 160 MeV/$c$ whereas the experimentally suggested 
value is about 1.3 and at a slightly higher $p_{\rm lab}$ of 165 MeV/$c$. Moving to higher $p_{\rm lab}$ $\simeq$ 
400 MeV/$c$, this ratio is about 2.3. It should be noted that in this low-energy region 
we have assumed that most of the $nA$ non-elastic reaction are due to the absorption process. We 
also restrict our analysis to the lowest momentum of 100 MeV/$c$ to 
avoid any complications due to contributions from the low-energy resonances.  

Again, as illustrated in Fig.~\ref{fig3}(a), the $\sigma^{\bar n \rm Fe}_{\rm ann}$/$\sigma^{ n \rm Fe}_{\rm rec}$ 
ratio is also turned out to be about 1.4 to 1.6 between $p_{\rm lab}$ values of 120 and 400 MeV/$c$. 
For the rest of the targets shown in Fig. \ref{fig4}(b), one finds that the $\sigma^{\bar nA}_{\rm ann}/\sigma^{ nA}_{rec}$ 
ratios vary between the order of 1.5 and 3.8 in the region where comparisons are possible, 
and also depend on both the momentum and the $A$ values. Notice that their momentum dependency of cross section ratios 
resembles their cross section behaviors, which are also quite different from those of the iron and carbon nuclei seen earlier in Fig.~\ref{fig3}(a).
Comparing to the case of carbon nuclei, Fig.~\ref{fig4}(b) indicates a much better agreement between the 
predicted and the experimental $\sigma^{\bar nA}_{\rm ann}$/$\sigma^{nA}_{\rm rec}$ ratios for all targets. 
The better agreement is understandable since the theoretical and experimental $\sigma^{nA}_{\rm rec}$ 
are also in a much closer agreement (e.g., see Fig.~\ref{fig7}). 

\begin{figure}[!htp]
\begin{center}
\includegraphics[scale=0.4]{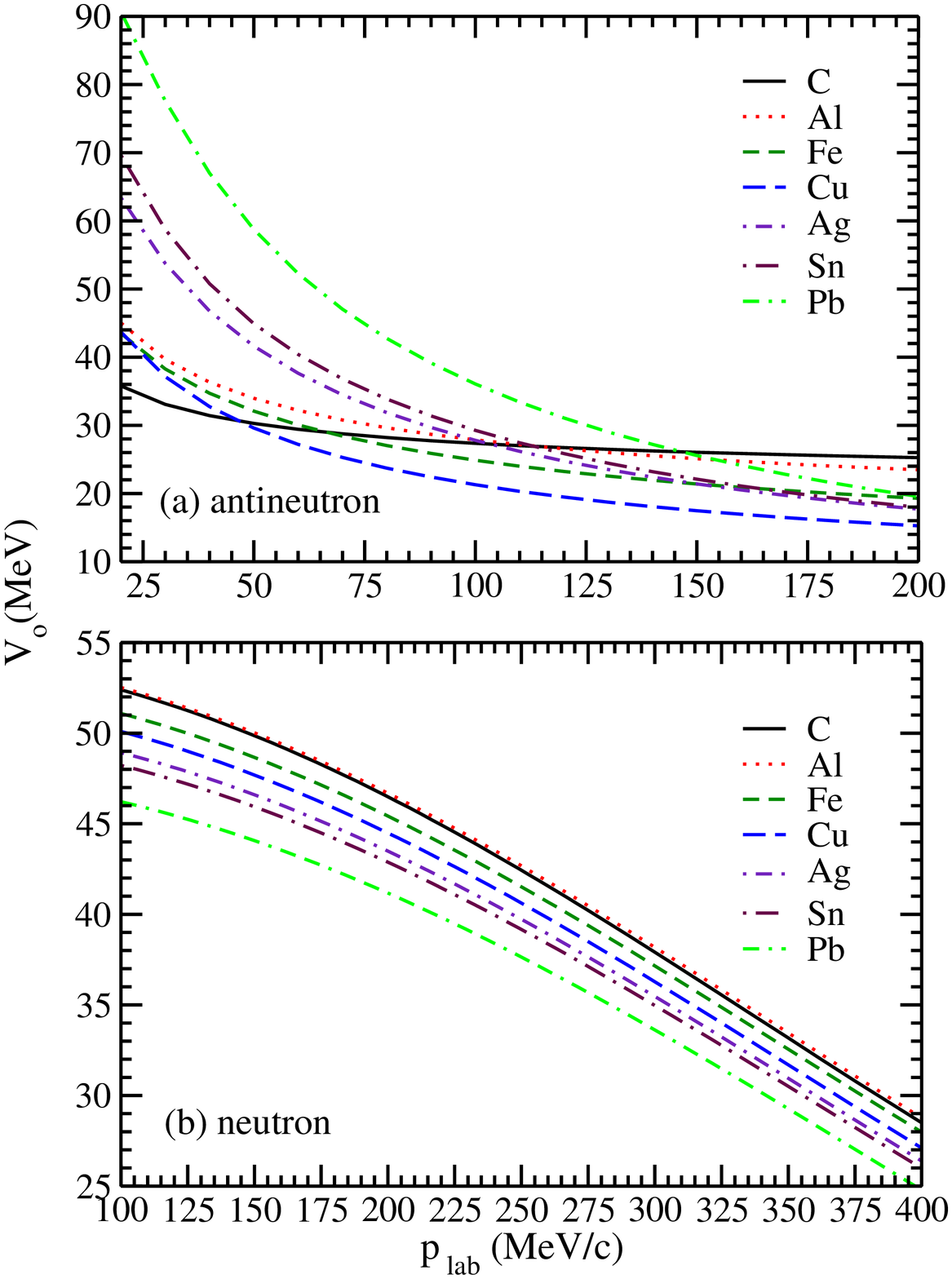}
\caption{(Color online) The variation of $V_o$ as a function of projectile momentum and atomic mass.}
\label{fig8}
\end{center}
\end{figure}

Now we consider the optical potential parameters for both $\bar nA$ and $nA$ interactions.
The values of the initial (or starting) potential depth $V'_o$ for all the target elements are given in Table \ref{t1}.
The $V'_o$ value, in general, increases from 52 to 110 MeV as the $A$ value goes from 12 to 206. 
But with KD-OMP \cite{Koning03} calculations, this trend is reversed for the case of the $nA$ reaction. 

\begin{figure}[!htp]
\begin{center}
\includegraphics[scale=0.4]{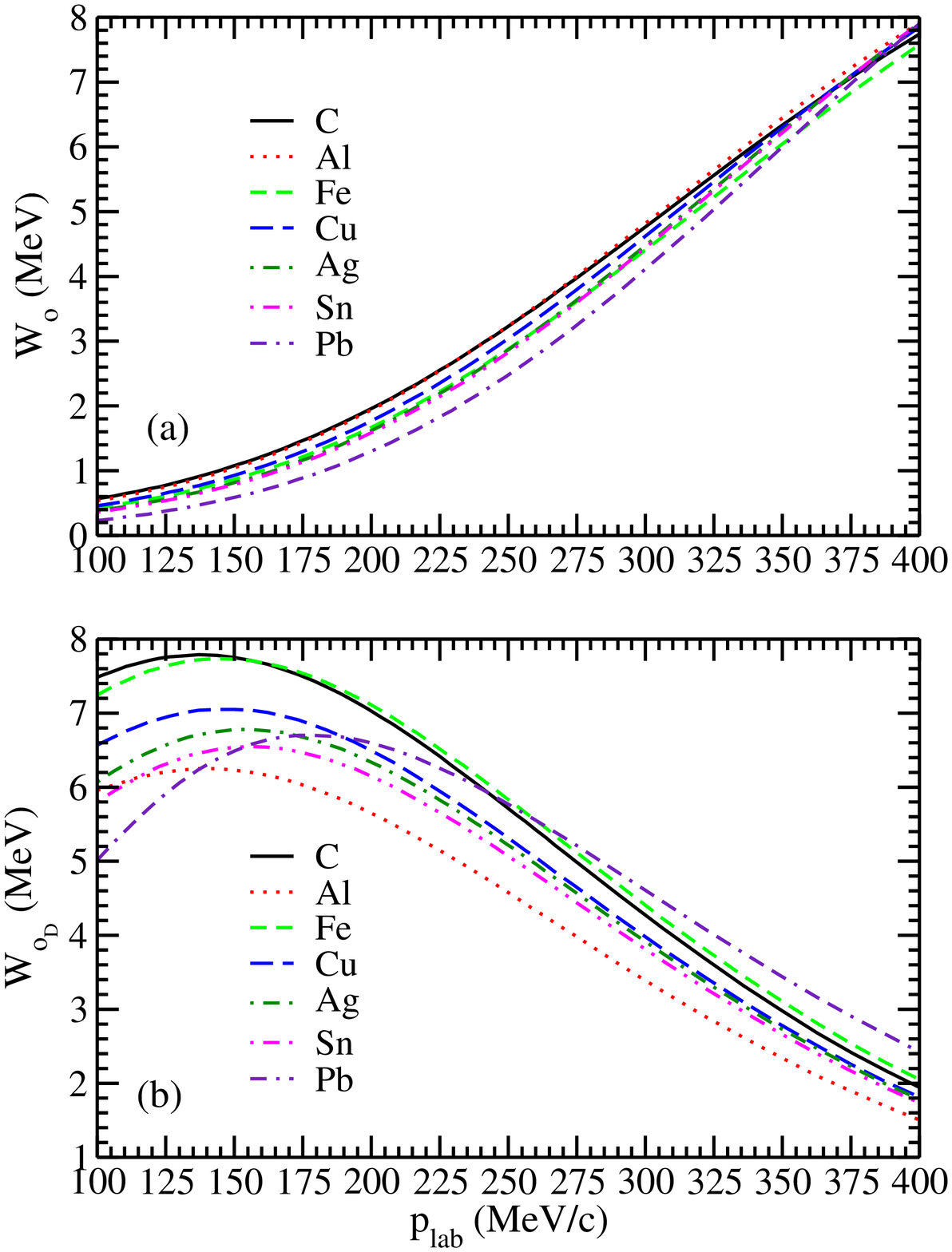}
\caption{(Color online) Neutron optical model potential well depth, $W_{(o,o_D)}$, as a function of projectile momentum.}
\label{fig9}
\end{center}
\end{figure}

The corresponding real parts of the central potentials $V_o$ for $\bar nA$ and $nA$ interactions, as a function of momentum, 
are shown in Figs.~\ref{fig8}(a) and \ref{fig8}(b), respectively. Although the depth of $V_o$ based on POMP for every nuclei 
decreases with increasing momentum according to Eq.(7), as shown in Fig.~\ref{fig8}(a), the antineutron's potential curves 
do not display any form of systematic order as a function of mass number $A$. At larger momentum (i.e., $p_{\rm lab}>$ 100 MeV/$c$ ), 
the potentials gradually become less sensitive to the increment of the projectile momentum.  
In contrast, in Fig.~\ref{fig8}(b), the neutron's $V_o$ obtained from KD-OMP  \cite{Koning03} for each nucleus does show 
a systematic decrease as the nuclear size increases and an almost linear decrease as a function of momentum, 
especially for $p_{\rm lab}>$ 200 MeV/$c$. 

The imaginary terms, W$_o$ and W$_{o_D}$, the volume and surface absorption POMP components, 
are also quite different from the KD-OMP prescribed values. First of all, they do not depend 
on projectile momentum. Second, as shown in Table~\ref{t2}, even though our W$_o$ for $\bar nA$ 
varies from 12.0 to 2.8 MeV with respect to carbon and to lead nuclei, 
there is no systematic change in W$_o$ as the nuclear size increases. In comparison to the case of $nA$, 
Fig.~\ref{fig9}(a) shows that the KD-OMP determined W$_o$ decreases as $A$ value increases, 
but increases as $p_{\rm lab}$ increases. Third, the antineutron's surface absorption W$_{o_D}$
 for $\bar n A$ is chosen to be a constant of 5.98 MeV for all targets. However, 
the neutron's surface absorption values W$_{oD}$ do depend on momentum and their functional forms are 
displayed in Fig.~\ref{fig9}(b). It should be noted that for neutrons, at low incident energy,
the absorption is dominated by the surface component W$_{o_D}$. Beyond about 250 MeV/$c$,
the volume term W$_o$ can no longer be ignored, and at higher energies the absorption
can be completely dominated by W$_o$.  

We compare the geometrical parameters $r_x$ and diffusiveness parameters
$a_x$ for $\bar nA$ and  $nA$ interactions in Table \ref{t3}. Similar to $nA$ interactions, 
with the case of iron nuclei as an exception, we have in the case of $\bar nA$ that the radii $r_W$ = $r_V$ 
and they increase as $A$ increases. But the present $r_V$ values for the antineutron
are significantly larger than those for the neutron. For example, the $r_V$ of 1.785 fm for $\bar n$Pb annihilation
is about 45 $\%$ larger than the $r_V$ of 1.235 fm for the $n$Pb reaction. Also, in the $\bar nA$ case, even though 
the $r_{W_D}$ values for $\bar nA$ and $nA$ are not that different, we have a constant value 
of $r_{W_D}$ = 1.26 fm for every nuclei, whereas the $r_{W_D}$ associated with  the $nA$ reaction 
decreases from the C target with $r_{W_D}$ = 1.306 fm to Pb with $r_{W_D}$ = 1.249 fm. 
A similar pattern is also found with the $\bar nA$ diffusiveness parameters $a_W$ = $a_V$ 
and $a_{W_D}$. The diffusiveness parameters $a_V$ for $\bar nA$ also happen to be at 
least a factor of 2-3 larger than those for the $nA$ interactions. Nevertheless, this set of POMP 
parameters enables us to obtain theoretical cross sections that complement the experimental 
annihilation cross sections across a wide momentum range. 

\subsection{$\bar pA$ annihilation cross sections}

As an adjunct to predicting the $\bar nA$ annihilation and $nA$ reaction cross sections, 
we further predict the $\bar pA$ annihilation cross section. We base our prediction on the 
simplest assumption that both $\bar pA$  and $\bar nA$ interactions 
have the same nuclear optical model potential but differs only in the long-range Coulomb interaction. 
The goal here is to examine the dependence of the annihilation cross sections on the projectile 
charge and to provide a benchmark for comparison against which the $\bar nA$ and 
$\bar pA$ interaction potentials may differ. 

In comparison to the neutral $\bar n$ projectile, according to the annihilation cross sections 
depicted in figures \ref{fig1}, \ref{fig2} and \ref{fig4}(a), it is within our expectation that 
the charged $\bar p$ projectile shows relatively larger annihilation cross section. As a matter of fact, 
because of the additional effects from Coulomb focusing, the $\bar p$ annihilation cross 
sections for all the nuclei feature a steeper rise than that of the $\bar nA$ interaction as the 
projectile momentum goes down. As the projectile momentum continues to increase, 
the effects from Coulomb focusing also gradually diminish. As a result, the annihilation 
cross sections for both $\bar n$ and $\bar p$ merge 
at $p_{\rm lab}$ $\sim$ 500 MeV/$c$, and eventually reaches Pomaranchuk's equality, 
in which their cross section ratio becomes unity at $\sim$1.0 GeV/$c$. These plots also evidently  
indicate that the $\bar pA$ annihilation cross sections are sensitive to the target mass number $A$.

To better understand the differences in annihilation cross sections due to $\bar p$ and $\bar n$ projectiles, 
we examine the $\sigma^{\bar pA}_{\rm ann}$/$\sigma^{\bar nA}_{\rm ann}$ ratios
as a function of momentum for carbon and iron nuclei in Fig. \ref{fig3}(b). The plots show that 
their behavior is similar to the momentum dependence of their annihilation cross sections, and
their slopes are remarkably steep in the region where the momentum goes to zero. 
Comparing the magnitude of the iron's ratio curve to that of carbon, one clearly sees a stronger Coulomb 
focusing effects for the heavier nucleus and this long-range effect weakens in the limit of large momemtum. 
In addition to that, Fig.~\ref{fig3}(b) also reveals a contrasting energy-dependent in the behavior 
of $\sigma^{\bar pA}_{\rm ann}$/$\sigma^{\bar nA}_{\rm ann}$ ratios in comparison 
to those of $\sigma^{\bar nA}_{\rm ann}$/$\sigma^{nA}_{\rm rec}$ ratios shown in Fig.~\ref{fig3}(a) . 

Fig.~\ref{fig4}(b) displays a collection of the behaviors of 
$\sigma^{\bar pA}_{\rm ann}$/$\sigma^{\bar nA}_{\rm ann}$ ratios
for all target nuclei as a function of momentum. The featured behavior is consistent 
with $\sigma^{\bar nA}_{\rm ann}$/$\sigma^{nA}_{\rm rec}$ 
where comparisons are possible, expected that the $\sigma^{\bar pA}_{\rm ann}$/$\sigma^{\bar nA}_{\rm ann}$ 
ratios are smaller by roughly a factor of 2. Again, all the $\sigma^{\bar pA}_{\rm ann}$/$\sigma^{\bar nA}_{\rm ann}$ 
ratios show strong momentum dependence at low momenta. 

Recently, the ASACUSA's Collaboration took a new measurement of the $\bar p$C 
annihilation cross section at a low energy of 5.3 MeV or $p_{\rm lab} =$ 100 MeV/$c$ 
\cite{asacusa18}. Their cross section value of 1.73 $\pm$ 0.25 
barns is also plotted in Fig.~\ref{fig1}. The datum clearly touches our prediction. In addition to that,
we have also plotted the one and 
only experimental datum for $\bar p$Sn at 100 MeV/$c$ in Fig.~\ref{fig4}(a4). 
The down side of this case are that there is no other comparable 
experimental measurements for $\bar p$ and $\bar n$ as in the 
case of protons. Therefore, at this point, we will not surmise 
the energy dependence of the $\bar p$Sn 
cross section.     

\subsection{The power laws and annihilation cross sections}

\begin{figure}[!htp]
\begin{center}
\includegraphics[scale=0.35]{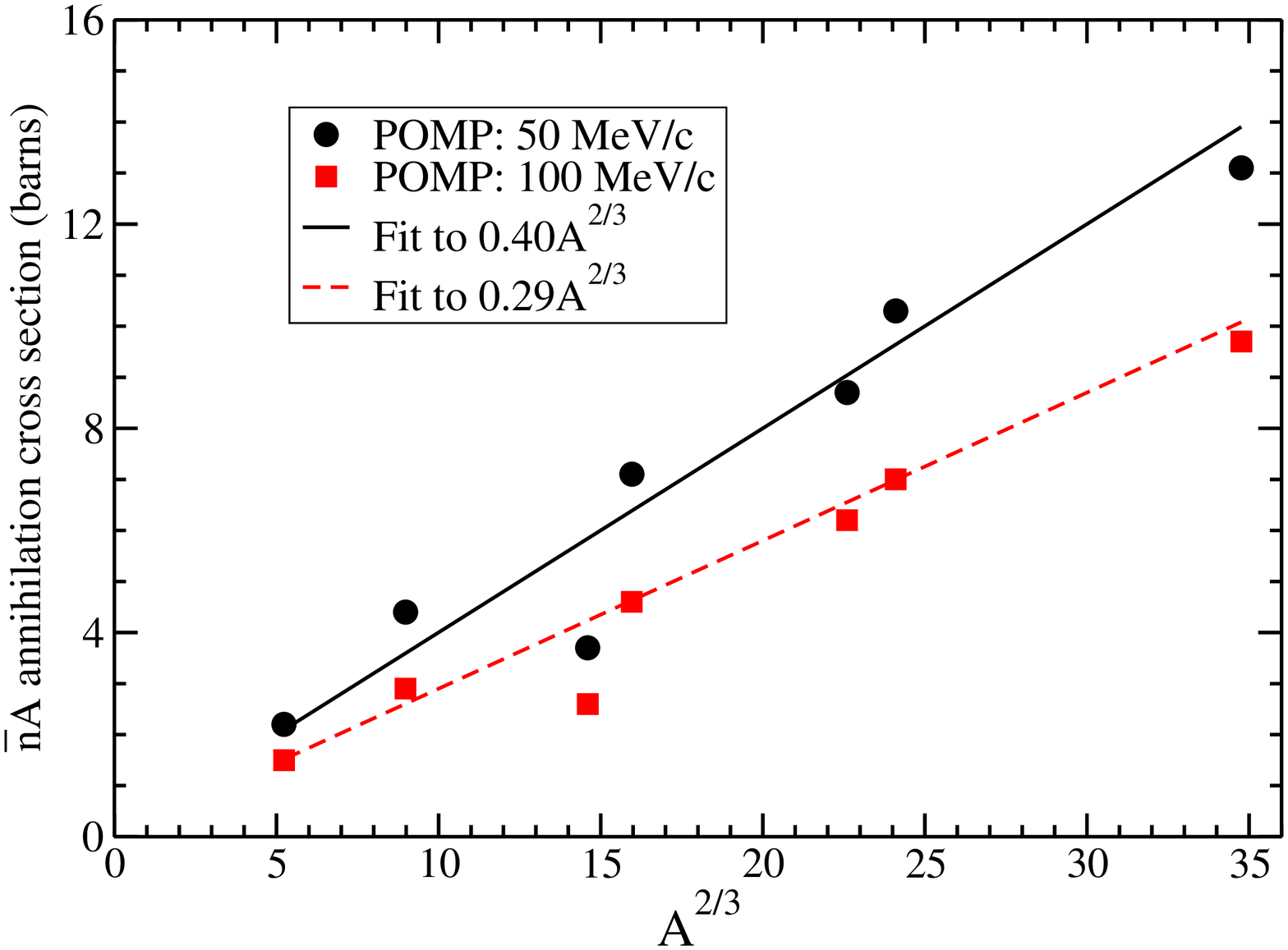}
\caption{(Color online) Antineutron annihilation cross sections as a function of atomic 
mass number $A^{2/3}$ at $p_{{\rm lab}}$ = 50 and 100 MeV/$c$. 
The scattered points are results from POMP calculations. 
The solid and dashed lines are results from 
the fitting to the expression $\sigma^{\bar nA}_{\rm ann}$ = $\sigma^{\bar nA}_o A^{2/3}$.}
\label{fig5}
\end{center}
\end{figure}  

Since it is of interest to find out whether $\sigma^{\bar nA}_{\rm ann}$ $\propto$ $A^{2/3}$ at low energies, 
we plotted $\sigma^{\bar nA}_{\rm ann}$ at $p_{\rm lab}$ = 50 and 100 MeV/$c$ against the corresponding 
mass number of $A^{2/3}$ in Fig. \ref{fig5}. The scattered points are the POMP predicted results. 
They are fitted with an expression of $\sigma^{\bar nA}_{\rm ann} = \sigma^{\bar nA}_o A^{2/3}$. The fitting is rather good. 
It indeed indicates that $\sigma^{\bar nA}_{\rm ann}$ has a linear dependence on $A^{2/3}$ at low energies. Apart from these, 
Fig.~\ref{fig5} additionally reveals that the $\bar n$Fe annihilation cross sections appear to peculiarly deviate 
from this linear dependence. Perhaps future experiments can reinvestigate this anomaly in the low-momentum region where $p_{\rm lab}$ 
is less than 100 MeV/$c$.  

\begin{figure}[!htp]
\begin{center}
\includegraphics[scale=0.325]{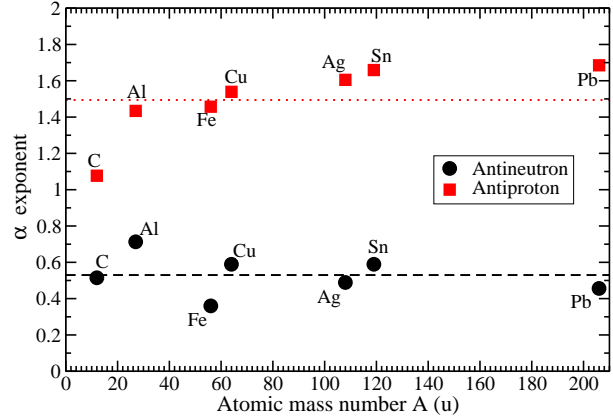}
\caption{(Color online) The exponent $\alpha$ expressing the dependence of 
$\sigma^{\bar nA}_{\rm ann}$, and $\sigma^{\bar pA}_{\rm ann} $ on $p_{\rm lab}$ 
as $\sigma^{\bar nA}_{\rm ann}$ $\propto$ 1/p$^{\alpha}_{\rm lab}$, as a function of the target mass number $A$. 
The scattered points are results from the present POMP calculations. 
The dashed line marks $\alpha$ = 0.530 (an average over all the targets) for the $\bar n$ projectile whereas the dotted line marks 
$\alpha$ = 1.494 (an averaged over all the targets) for the $\bar p$ projectile.}
\label{fig6}
\end{center}
\end{figure}  

It is also informative to examine the inverse power law of $\bar nA$ annihilation.
In the limit of low-energy, parametrizing the theoretical annihilation cross section 
in an inverse power law form, $\sigma^{\bar nA}_{\rm ann}$ $\propto$ 1/p$^{\alpha}_{\rm lab}$,  
in the range between 40 and 100 MeV/$c$, the $\alpha$ exponential value can 
be easily determined by setting $\alpha = \partial \ln(\sigma_{\rm ann})/\partial  \ln(p_{{\rm lab}})$. 
Fig.~\ref{fig6} gives the variation of $\alpha$ exponential values as a function of mass number $A^{2/3}$. 
Taking an average over all the nuclear targets yields a value of $\alpha = 0.530$. This 
consequently suggests that the $\sigma^{\bar nA}_{\rm ann}$ may be proportional to 1/$p^{1/2}_{{\rm lab}}$ 
for targets with $A~\ge$ 6. This finding appears to be far from what we learned in our 
previous work \cite{Lee16}. There we found in the case of $\bar n p$, the exponential 
value $\alpha$ = 1.08 in the momentum range between 30 and 95 MeV/$c$. This exponential 
value is very close to the expected $\alpha$ = 1.0 value, a clear indication of the $1/p_{\rm lab}$ 
behavior. However, in our previous study \cite{Lee16}, the nuclear potential was assumed to be a constant there. 
Here, in contrast, the nuclear optical potential depends on the projectile momentum, 
causing the $\sigma_{\bar n A}$ to deviate from the $1/p_{\rm lab}$ law. 

At the low-energy limit, we can see that the cross section slope for the $\bar p A$ interaction is much 
steeper than the one of $\bar n A$. Therefore, it is also meaningful to check the inverse 
power law form, $\sigma_{\rm ann}^{\bar p A}$  $\propto$ 1/p$^{\alpha}_{\rm lab}$, of $\bar pA$ annihilation. 
Similar to what we have discussed earlier with respect to $\bar nA$ annihilation in Fig. \ref{fig6},
parametrizing the theoretical  annihilation cross section in a power law form in the range between 
40 and 100 MeV/$c$ allows one to obtain the $\alpha$ exponential value. In our previous investigation 
on $\bar p p$ interaction \cite{Lee16} , we found that $\alpha$ = 1.544 in the momentum range between 30 and 50 MeV/$c$. 
Displayed in Fig. \ref{fig6} is the variation of $\alpha$ as a function of mass $A$. Similarly, averaging 
these values over the seven nuclear targets yields a value of $\alpha$ = 1.494. As opposed to the case of $\bar n A$, 
this value is close to what we found previously in the case of $\bar p p$ annihilation. This also means the 
Coulomb effect is dominant at the low-energy limit and cannot be neglected. The extracted $\alpha$ = 1.494 is not quite 
equal to $\alpha$ = 2.0 as expected at the very-low-energy limit \cite{Wig48,Lan58}.
This means that the approach to the lowest energy limit of $\alpha=2$ will occur at much lower energies
than the range of low energies considered here.

\section{Summary and Conclusions}
The purpose of this contribution is two-fold. The first one is to revisit and rectify our 
previous annihilation cross section results for $\bar nA$ in \cite{Lee16}. The second one is 
to pursue a phenomenological analysis of $\bar n$ annihilation cross section 
as a function of projectile momentum $p_{\rm lab}$ and mass number $A$.  

Previously, we used the extended Glauber theory \cite{Lee16} 
to examine the experimental annihilation cross section data for $\bar n$ 
on C, Al, Fe, Cu, Ag, Sn, and Pb in the momentum range 
below 500 MeV/$c$.  But an inadvertent
error arose through the Coulomb trajectory modification, 
causing the results to agree with the experimental data. After amending
the theory, the re-evaluated results turned out to be in disagreement with the experimental data.

The Glauber theory is well known to be valid for high-energy collisions 
in which the extend individual nucleon can be treated as an isolated scatterer.
For low-energy collisions, such description may not be as appropriate 
and the traditional optical model analysis may be more suitable. For this reason 
we adopt the optical model potential to analyze the momentum dependence 
of $\bar n A$ annihilation cross section. 

The use of a microscopic optical model potential method was previously attempted 
by Friedman \cite{Fri14, Fri15} to investigate the momentum dependence of $\bar n A$ annihilation cross sections. 
The investigation found that the annihilation cross section of $\bar n$ on nuclei cannot be described 
by a microscopic optical potential that fits well the available data on the $\bar p$ interactions with nuclei. 
Nevertheless, inspired by the works of Friedman and Koning and Delaroche \cite{Koning03}, 
we explored a new form of momentum dependent optical model potential to 
describe the $\bar nA$ interaction. Even though it is phenomenological and local, 
the presented optical model potential of Eq.(7) is quite different 
from that of the Koning and Delaroche and that of Friedman. It is simple, as well as 
comprehensive enough to treat very-low-momentum $\bar nA$ 
and $\bar pA$ annihilations. We employed the momentum-dependent optical model potential 
in the Schr\"{o}dinger equation and the equation is solved using 
the standard distorted wave method provided in the ECIS97 computer 
program \cite{Raynal97} to evaluate the annihilation cross sections for $\bar nA$ and $\bar pA$.
Similarly, we have also applied the Koning-Delaroche's momentum-dependent 
optical model potential to examine the $nA$ non-elastic reaction 
cross sections on on C, Al, Fe, Cu, Ag, Sn, and Pb. We showed that the 
calculated cross sections are in reasonable agreement with the recommended 
data from Brookhaven National Laboratory's database. 
  
Although, in this study, we found that the present $\bar nA$ annihilation cross sections 
fit the experimental data rather well, this does not mean that we have
fundamentally understood the neutral $\bar nA$ annihilation 
mechanism. In fact, the opposite is true. 
For a start, even though both the $\bar nA$ and $nA$ interactions are Coulomb-free, 
why does the $\sigma^{\bar nA}_{\rm ann}/\sigma^{ nA}_{\rm rec}$ 
cross section ratio appears to be so large (almost by a factor of 2)? From a simple 
geometrical argument,  in comparison to the 
incoming neutron $n$, why does the antineutron $\bar n$ seems to have a 
larger ``effective area" for the target nuclei to react? Further theoretical and 
experimental efforts are necessary to address these fundamental questions.

In the low-energy range considered here, we have demonstrated and verified that $\sigma^{\bar nA}_{\rm ann}$ 
is indeed approximately proportional to $A^{2/3}$. We have illustrated that for neutral or 
Coulomb-free $\bar n A$ interactions the annihilation $\sigma^{\bar nA}_{\rm ann}$ $\propto$ 1/$p_{\rm lab}^\alpha$.
In addition, we have also shown that the $\alpha$ value for charged $\bar p A$ interactions is significantly larger 
than the $\alpha$ value for the neutral $\bar nA$ interactions. We presume that this is likely 
due to the additional Coulomb effects on top of nuclear interactions for charged $\bar p A$ interactions. 
In conclusion, we have calculated the $\bar n A$ annihilation cross section based 
on the simplest assumption that both $\bar n A$ and $\bar pA$ interactions 
have the same nuclear optical potential but differ only in the long-range 
electrostatic interaction. Any deviation from such a simple model extrapolation 
in measurements will shed new and desirable information on the difference between $\bar nA$ 
and $\bar pA$ potentials. 

\vspace*{0.1cm}
\centerline {\bf Acknowledgment}
\vspace*{0.2cm} 
The authors thank Drs.~Robert Varner, Yuri Kamyshkov and Luca Venturelli for helpful 
discussions and communications. The research was supported in 
part by the Division of Nuclear Physics, U.S. Department of Energy 
under Contract No. DE-AC05-00OR22725.

\end{document}